\documentclass[a4paper,12pt]{article}
\usepackage[utf8]{inputenc}
\usepackage[T1]{fontenc}
\usepackage{amsthm,color, graphicx, amsmath}
\usepackage{bm}

\usepackage[retainorgcmds]{IEEEtrantools}
\usepackage{epstopdf}

\begin{document}

\title{ Influence of Initial Residual Stress on  Growth and Pattern Creation for a Layered Aorta}

\author{Yangkun Du$^{1,5}$, Chaofeng L\"{u}$^{2,3}$, Michel Destrade$^{1,5}$, Weiqiu Chen$^{1,3,4}$\\[12pt]
Department of Engineering Mechanics, \\ Zhejiang University, Hangzhou, P. R. China;\\[6pt]
Department of Civil Engineering, \\ Zhejiang University,  Hangzhou 310058, P.R. China; \\[6pt]
Soft Matter Research Center, \\  Zhejiang University, Hangzhou 310027, P. R. China; \\[6pt]
Key Lab of Soft Machines and Smart Devices of Zhejiang Province, \\ Zhejiang University, Hangzhou, P.R. China; \\[6pt]
Stokes Centre for Applied Mathematics,\\ School of Mathematics, Statistics and Applied Mathematics,\\ NUI Galway, Galway, Ireland}

\date{}

\maketitle


\begin{abstract}


Residual stress is ubiquitous and indispensable in most biological and artificial materials, where it sustains and optimizes many biological and functional mechanisms. \color{black}
The theory of volume growth, starting from a stress-free initial state, is widely used to explain the creation and evolution of growth-induced residual stress and the resulting changes in shape, and to model how growing bio-tissues such as arteries and solid tumors develop a strategy of pattern creation according to geometrical and material parameters.\color{black}
This modelling provides promising avenues for designing and directing some appropriate morphology of a given tissue or organ and achieve some targeted biomedical function. 
In this paper, we rely on a modified, augmented theory to reveal  how we can obtain  growth-induced residual stress and pattern evolution of a layered artery by starting  from an existing, non-zero initial residual stress state. 
We use experimentally determined residual stress distributions of  aged bi-layered human aortas and quantify their influence  by a magnitude factor.  
Our results show that initial residual stress has a \color{black} more \color{black} significant impact on residual stress accumulation and the subsequent evolution of patterns \color{black} than geometry and material parameters.\color{black}
Additionally, we provide an essential explanation for growth-induced patterns driven by differential growth coupled to an initial residual stress. 
Finally, we show that initial residual stress is a readily available way to control growth-induced pattern creation for tissues and thus \color{black} may provide \color{black} a promising inspiration for biomedical engineering.


\end{abstract}


\section*{Introduction}


Many bio-tissues, such as arteries, heart, brain, intestine and some tumors, are under significant levels of residual stresses in vivo and also once unloaded\cite{Holzapfel2007, Holzapfel2010, Omens1990, Savin2011, Balbi2015, Stylianopoulos2012, Fernandez-Sanchez2015}.
Residual stresses are used for maintaining a self-balanced state, by transferring physical signals and regulating some specific bio-func\-tions\cite{Helmlinger1997,Padera2004,destrade2012uniform,Pan2016}.  
It is fair to say that healthy biological performances rely heavily on appropriate levels of residual stress.
Hence, a greater  understanding of the role played by residual stress can lead to better design of mechanical, electrical, chemical, biological and internal environments for living organisms. 

In biology, residual stress is widely accepted as the result of growth and remodelling processes or of other, more involved bio-interactions. 
Physically, by referring to the multiplicative decomposition (MD) method of plasticity theory,  and by decomposing the overall growth process into two separate parts, Rodriguez et al. \cite{Rodriguez1994} were able to simulate the growth process and to explain growth-induced residual stress. 
The former step refers to the mass accumulation and incompatibility creation processes within two stress-free states called unconstrained growth; the later step refers to residual stress creation and compatibility restoration processes called elastic deformation.
This MD model illuminates many related mathematical and mechanical studies on the growth of artery, brain, intestine, etc.\cite{ Cao2012, Li2011, Lu2016, Du2017, Balbi2015, Wang2017}. 
Moreover, the influence of growth factors, including differential growth extent,  volume change rate, and/or growth velocity, on residual stress distribution, pattern creation and evolution has also been analysed with the MD model\cite{Goriely2005, Ciarletta2013, Ciarletta2014, Balbi2015}.

\begin{figure}
	\centering
	\includegraphics[width=0.95\textwidth]{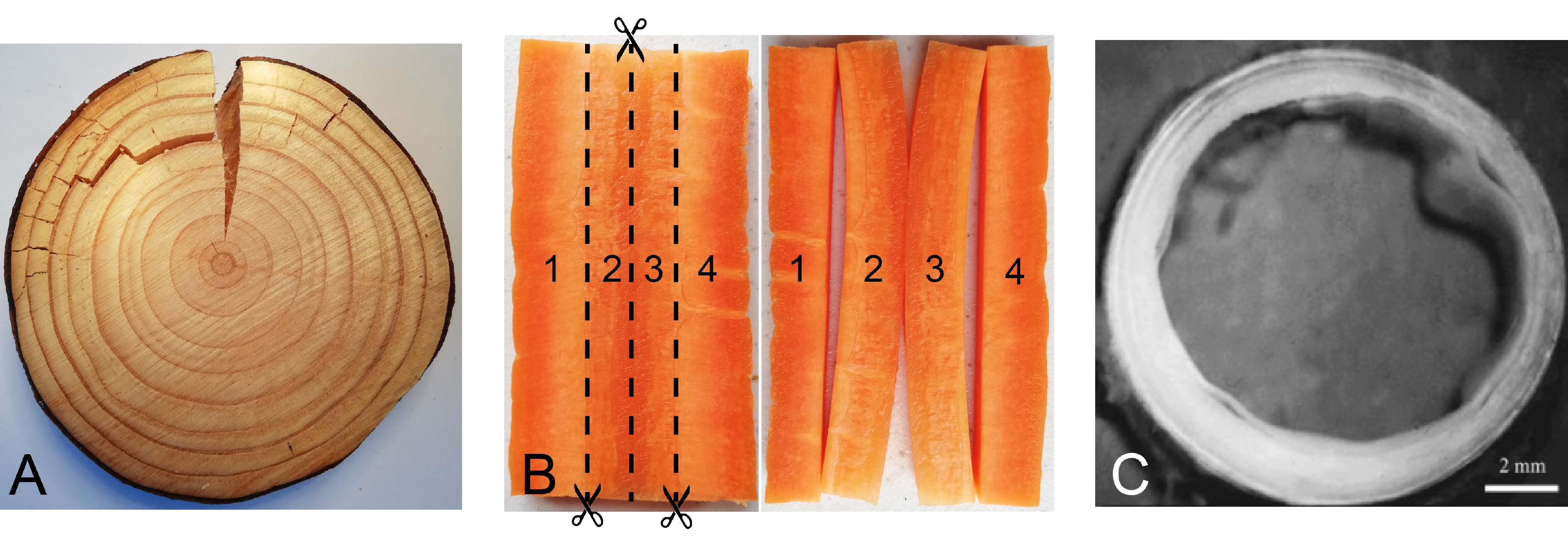}
	\caption{(A) Cutting radially through a slice of Irish Ash tree releases what were clearly high levels of residual stresses in the circumferential and radial directions;
	 (B) Cutting an axial slab of fresh carrot along the dashed lines reveals that the outer layers (1 and 4) were almost stress-free whilst the core layers (2 and 3) were subjected to large inhomogeneous axial and radial  residual stresses; (C) Layer-specific structure of a cut human aorta ring (from Holzapfel et al.\cite{Holzapfel2007}), revealed once it is unloaded from blood pressure and from axial and radial residual stresses: the circumferential residual stress causes a consequent contraction in the unloaded state, leading to buckling and even delimitation on the inner surface.}
	\label{Liver}
\end{figure}

Nonetheless, there  still remain some limitations for this  model due to its assumption that both the initial configuration and the natural configuration (or virtual configuration \cite{Johnson1995}) remain stress-free states during the unconstrained growth process.
As discussed by Johnson et al.\cite{Johnson1995}, for residually stressed materials  the stress-free state is an entirely discrete state made of  a collection of nearly infinitesimal volumes, which is unattainable in practice for bio-tissues. 
As shown by Figure \ref{Liver}(A) and (B), the residual stress in biological tissues can  be released only partially by cutting them in different directions.
A complete release of residual stress would require an infinite number of cuts.

Similarly, some investigations on stress-dependent or strain-dependent growth law cannot be set up properly  without using the actual residual stress.
By dividing the total growth deformation into many small steps that are further decomposed by the MD model, Goriely and Ben Amar \cite{Goriely2007} proposed an incremental, cumulative, and continuous growth law, allowing some growth steps to develop from a residually stressed state. 
With the exception of the very first step, every step grows from a residually stressed state, which provides a way to analyze the influence of residual stress on the growth process. 
However, the residual stress in any accumulative growth step stems from the previous growth step; most importantly, it is still necessary to assume an initial residual stress-free state in the first step, something which, again, is very difficult to prescribe and measure in real biological tissues.

It follows that  the MD growth model must be modified to account for the existing initial residual stresses found in living, growing continuum tissues, which result from complicated growth and remodelling processes or other bio-interactions \cite{Stylianopoulos2012, Ciarletta2016}.

Hence, taking the initial state to be an arbitrary residual stress state and building on the residual stress theory of Hoger et al.\cite{Hoger1986, Johnson1995, Shams2011, Gower2015}, we recently proposed\cite{Du2018} a modified multiplicative decomposition growth (MMDG) model. 
In this paper we set out to show the influence of initial residual stress on  growth-induced residual stress and the following pattern creation by focusing on a scenario modeling a three-dimensional, layered, initially stressed, growing aorta. 

We adopt some existing \color{black} experimental and theoretical \color{black} results of aortic residual stress distributions for our initial residual stress, and discuss its influence by \color{black} way \color{black} of a magnitude factor.
We compute the distributions of growth-induced residual stress and the geometrical changes of the aorta starting from different initial residual stress states. 
Based on a subsequent stability analysis for the growing aorta with initial residual stress, we obtain the critical wrinkling solution using linear incremental theory and surface impedance methods. 
We investigate the influence of the initial residual stress level and \color{black} of \color{black} the differential growth extent on pattern development for the layered aorta and provide sound explanations for their effect on pattern creation.

Finally, we emphasise that our results may provide an effective strategy to obtain  targeted pattern\color{black}s \color{black} by controlling the initial residual stress instead of the differential growth or swelling extent \cite{Ciarletta2014, Ciarletta2013}. 
Hence by going beyond bio-tissue growth, this work can also \color{black}provide \color{black} inspiration for industrial manufacturing, nano-fabrication, or self-assembly of soft solids which display behaviors similar to growth or swelling \cite{Yin2009, Chen2010, Chen2013, Li2013, Bertrand2016}. 


\section{MMDG framework and basic equations}



Based on the concept of multiplicative decomposition of  the total growth process, the overall growth process for an initially stressed material is also decomposed into separate parts. 
As shown in Figure 2, our innovation is to introduce an initial elastic deformation $\bm{F}_0$ for releasing the initial residual stress to a virtual stress-free configuration. 
Thereafter, the pure growth deformation $\bm{F}_g$ can happen between two virtual stress-free and incompatible configurations. 
Finally, a pure elastic deformation $\bm{F}_e$ is induced when the tissue restores the compatibility of the bio-tissue after growth, which \color{black} is the direct source of \color{black} the associated residual stress.

According to this new description of the growth process, the total growth deformation is decomposed as 
\begin{equation}
	\boldsymbol{F}=\boldsymbol{F}_e\boldsymbol{F}_g\boldsymbol{F}_0.
	\label{F}
\end{equation}

\begin{figure}[h]
	\centering
	\includegraphics[width=0.7\textwidth]{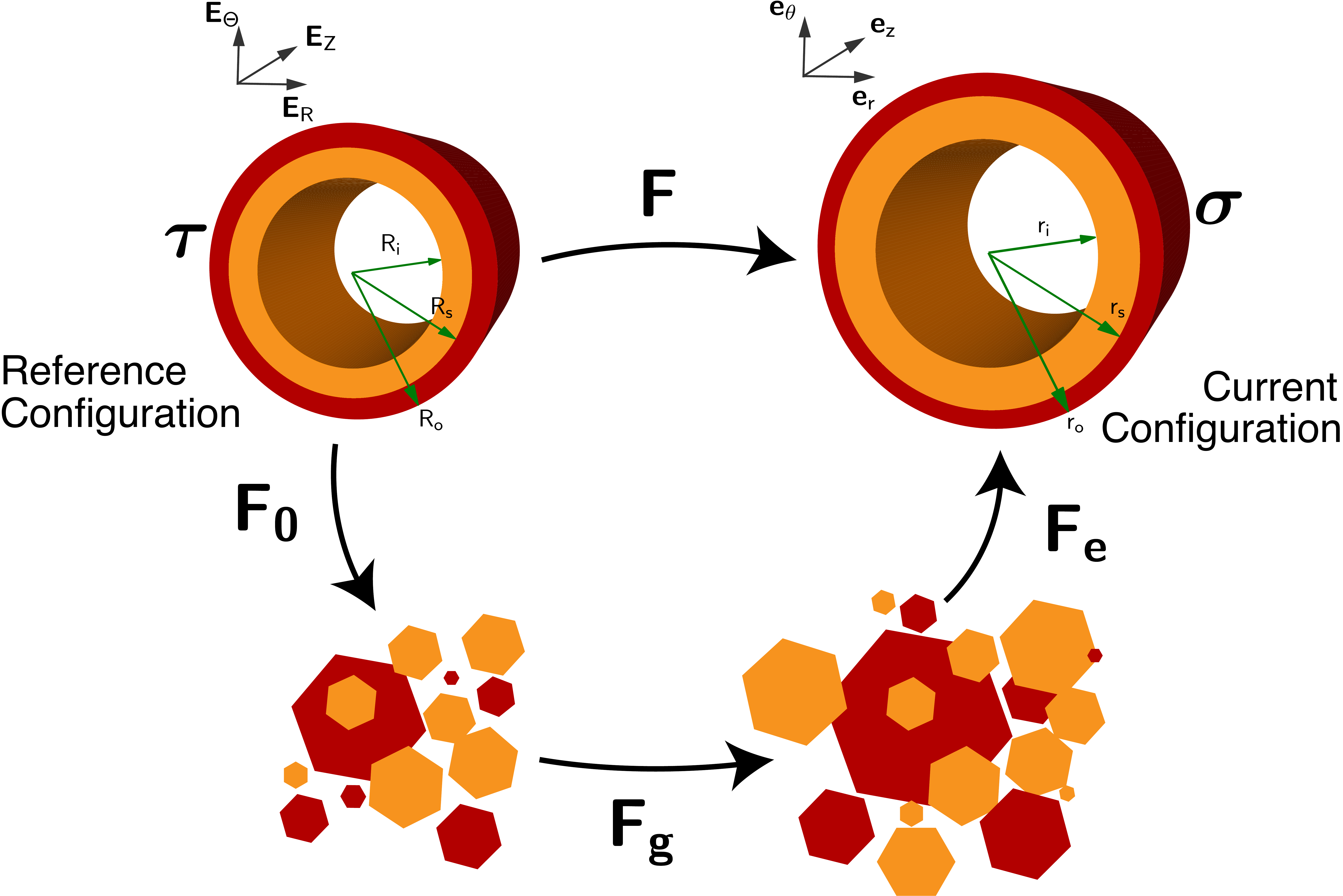}
	\caption{Modeling of the whole process of growing a bi-layered artery with initial residual stress. 
	Here the orange (inner) part is made of the combination of the intima and media layers, and the red  (outer) part is the adventitia layer.}
	\label{IS}
\end{figure}
Here we assume that the material in the virtual stress-free state is a natural material  following the constitutive equation of an incompressible neo-Hookean solid, also adhering to the initial stress symmetry condition \cite{Gower2015} although other models could equally have been chosen, e.g.\cite{Shams2011,merodio2013influence,ahamed2016modelling,merodio2016extension}. 
Hence, from Eq.\eqref{F} follows the constitutive equation for the Cauchy stress $\boldsymbol \sigma$ of growing bio-tissues with initial residual stress $\boldsymbol \tau$ as
\begin{equation}
	\boldsymbol{\sigma}=J_g^{-{2}/{3}} \left( \boldsymbol{F\tau F}^T+p_0\boldsymbol{FF}^T \right)-p\boldsymbol{I},
	\label{constitutive equation}
\end{equation}
where $p$,  $p_0$ are the Lagrange multipliers in the current and reference configurations, respectively. Here we assume  isotropic growth, so that $\boldsymbol{F}_g=J_g^{{1}/{3}} \boldsymbol{I} = \text{diag} \{g^k, g^k, g^k\}$, where $J_g=\det \boldsymbol{F}_g$ is the volume change ratio. 
The initial residual stress $\boldsymbol \tau$ and the Cauchy stress $\boldsymbol \sigma$ satisfy the equilibrium equations and boundary conditions
\begin{equation}
	\text{Div} \boldsymbol{\tau}=\boldsymbol{0},\qquad \boldsymbol{\tau N}=\boldsymbol{0},
	\qquad \text{div} \boldsymbol{\sigma}=\boldsymbol{0},\qquad \boldsymbol{\sigma n}=\boldsymbol{0},
	\label{equilibrium equation}
\end{equation}
where $\boldsymbol{N}$ and $\boldsymbol{n}$ are the normal vectors to the surfaces in the reference and current configurations, respectively.


\section{A growing aorta with an experimentally-de\-ter\-mi\-ned initial residual stress field}


Layered tubular tissues are the focus of this paper, as they are common in the body, e.g. artery, airway, intestine, etc. 
They are all known to possess a high degree of residual stress, as shown experimentally by cuts.
It is also known that their bio-functions depend significantly on their residual stress levels and their morphology.  
Figure \ref{Liver}(C) shows a ring of an aged human aorta, which can be modelled as a bi-layered tube  according to  Holzapfel and collaborators \cite{Holzapfel2007,Holzapfel2010}. 

Here we follow their model and take the aorta as a bi-layer, by combining the intima and the media into one inner layer and having the adventitia as the outer layer in a three-dimensional Euclidean space. 
The reference configuration with initial residual stress is associated with the cylindrical coordinates  $\left( R, \Theta, Z \right)$, and the current configuration with a new residual stress state is associated with the coordinates $\left(r, \theta, z \right)$. 

Holzapfel and collaborators \cite{Holzapfel2007, Holzapfel2010} determined the distribution of residual stress in aged human aortas from measurements of the circumferential opening angles and axial bending angles for each layer, see Figure \ref{IS}(A) for the protocol. 
We reproduced their results (details not shown here) to get $\boldsymbol \tau$ in non-dimensional form, see Figure \ref{IS}(B) (and compare with the dimensional version\cite{Holzapfel2010}). 
From here on, we multiply this distribution by a factor $\alpha$ to introduce a magnitude factor of the initial residual stress and clearly illustrate its effects: $\alpha=0$ corresponds to the stress-free state, $\alpha=1$ corresponds to the initial stress of Holzapfel and collaborators \cite{Holzapfel2010, Holzapfel2007}. 
\color{black} We take their shear moduli $\mu^{in}=34.4$ kPa for the inner layer and $\mu^{ad}=17.3$ kPa for the adventitia layer, respectively. \color{black}
In addition, for convenience and better display of the results, we scale down the dimensions of the bi-layered aorta by a factor 10, resulting in the following geometry: $R_i=0.5911$ mm, $R_s=0.6724$ mm, $R_o=0.7504$ mm, for the inner, interface, and outer radii, respectively. 

The radial stress is continuous across the aortic thickness and remains compressive throughout, while the circumferential and axial stresses are discontinuous at the interface between the layers. 
The circumferential stress changes from compressive on the inner surface of the intima to tensile at the interface, and remains tensile in the adventitial layer. 
The axial stress is compressive in the inner layer and tensile in the adventitial layer. 
The axial stress has the largest magnitude, and the circumferential stress is also consequent; the radial stress is almost negligible in comparison, see Figure \ref{IS}(B).

\begin{figure}[h]
	\centering
	\includegraphics[width=1\textwidth]{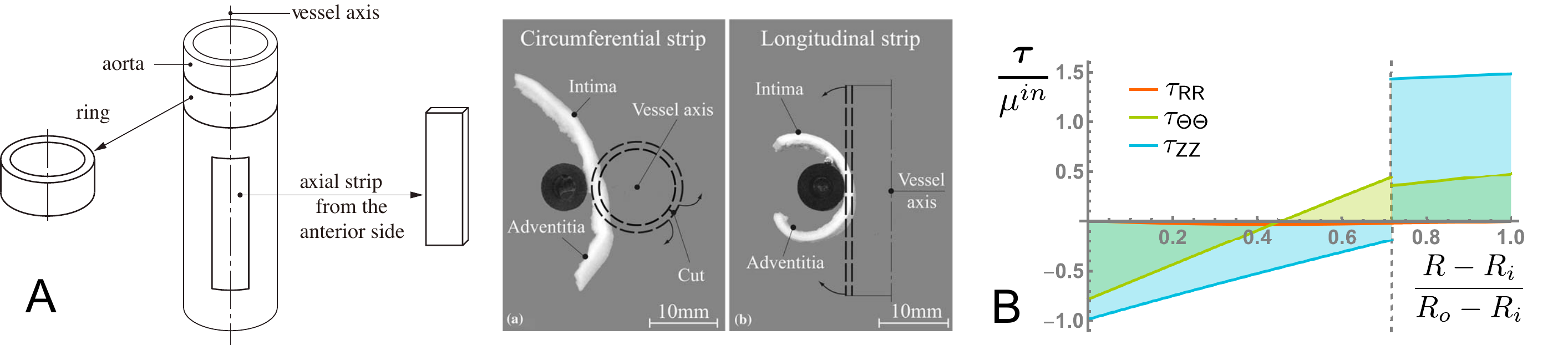}
	\caption{(A) Experimental protocol for measuring the three-dimensional layer-specific residual stress distribution of a human aorta (from Holzapfel et al.\cite{Holzapfel2007}); (B) The resulting non-dimensional transmural distributions of initial residual stresses in a bi-layered aged human aorta. In this paper, they are magnified by a factor $\alpha > 0$.}
	\label{IS}
\end{figure}

According to the above bi-layered modeling, we now impose the constrained growth deformation (overall growth deformation) gradient tensors in the residually stressed inner and adventitia layers by prescribing $\boldsymbol{F}^k= \text{diag}\left({\mathrm{d}r^k}/{\mathrm{d}R^k},{r^k}/{R^k},1\right)$, where $k=in, ad$ denotes the corresponding quantities in the inner layer and adventitia, respectively. 
Then, based on Eq. \eqref{constitutive equation}, we compute the following non-zero Cauchy stress components in the current state,
\begin{equation}
	\begin{split}
		&\sigma_{rr}^k=\left(J_g^k\right)^{-\frac{2}{3}}\left(\tau_{RR}^k+p_0^k\right)\left(\frac{J_g^k R^k}{r^k}\right)^2-p^k,\\
		&\sigma_{\theta\theta}^k=\left(J_g^k\right)^{-\frac{2}{3}}\left(\tau_{\Theta\Theta}^k+p_0^k\right)\left(\frac{r^k}{R^k}\right)^2-p^k,\\
		&\sigma_{zz}^k=\left(J_g^k\right)^{-\frac{2}{3}}\left(\tau_{ZZ}^k+p_0^k\right)-p^k,
	\end{split}
\end{equation}
where $p_0^k$, $p^k$ are yet undetermined variables: $p^k$ depends on the boundary conditions in the current configuration, and  $p_0^k$ is determined from the incompressibility condition for the natural materials,
\begin{equation}
	\det\left[\left(\boldsymbol{F}_0^k\right)^{-1}\left(\boldsymbol{F}_0^k\right)^{-T}\right]=\det\left(\frac{\boldsymbol{\tau}^k+p_0^k\boldsymbol{I}}{\mu^k}\right)=1.
\end{equation}
Now that we have the distribution of the initial residual stress,  $p_0^k$ is found in terms of $\boldsymbol{\tau}^k$ and shear moduli $\mu^k$ as $p_0^k=\frac{1}{3}\left(T_3^k+{T_1^k}/{T_3^k}-I_{1,\boldsymbol{\tau}^k}\right),$
where $T_1^k=I_{1,\boldsymbol{\tau}^k}^2-3 I_{2,\boldsymbol{\tau}^k}$, $T_3^k=\sqrt[3]{\sqrt{\left(T_2^k\right)^2-\left(T_1^k\right)^3}-T_2^k}$, $T_2^k=I_{1,\boldsymbol{\tau}^k}^3-\frac{9}{2} I_{1,\boldsymbol{\tau}^k}+\frac{27}{2}\left(I_{3,\boldsymbol{\tau}^k}-\left(\mu^k \right)^3\right)$,  and $I_{1,\boldsymbol{\tau}^k}$, $I_{2,\boldsymbol{\tau}^k}$, $I_{3,\boldsymbol{\tau}^k}$ are the principal invariants of the initial residual stress tensor.
Then, from the incompressibility condition we find that 
\begin{equation}
		r^{k}=\sqrt{J_g^{k}\left(\left(R^{k}\right)^2-(R_a^k)^2\right)+(r_a^k)^2},
	\label{geometric governing eq}
\end{equation}
where $a={i,s}$ corresponds to $k={in,ad}$ respectively. 
%

Then, based on the equilibrium equation Eq.\eqref{equilibrium equation}$_2$ and the boundary condition $\sigma_{rr}^{in}|_{r=r_i}=\sigma_{rr}^{ad}|_{r=r_o}=0$ in the current configuration, the radial Cauchy stress is obtained as
\begin{multline}
		\sigma_{rr}^{k}\left(\varsigma^{k}\right)=
		\int_{\varsigma_0^{k}}^{\varsigma^{k}} \left[\left(\tau_{\Theta\Theta}^{k}+p_0^{k}\right)\left(\frac{r^{k}}{R^{k}}\right)^2-\left(\tau_{RR}^{k}+p_0^{k}\right)\left(\frac{J_g^{k} R^{k}}{r^{k}}\right)^{2}\right] \\ \times \frac{\left(J_g^{k}\right)^{\frac{1}{3}}H^{k}\left(\varsigma^{k}H^{k}+R_a^k\right)}{J_g^{k} \varsigma^{k} H^{k}\left(\varsigma^{k}H^{k}+2R_a^k\right)+(r_a^k)^2}\mathrm{d}\varsigma^{k}\\
		\label{CS}
\end{multline}
where $H^{in}=R_s-R_i, \varsigma^{in}=\frac{R-R_i}{H^{in}},  \varsigma_0^{in}=0, H^{ad}=R_o-R_s, \varsigma^{ad}=\frac{R-R_s}{H^{ad}}, \varsigma_0^{ad}=1$.
Now the Cauchy stress profile is determined by combining  Eqs.\eqref{geometric governing eq}-\eqref{CS} and the continuity condition $\sigma_{rr}^{in}\left(1\right)=\sigma_{rr}^{ad}\left(0\right)$ at the interface.

\begin{figure*}
	\centering
	\includegraphics[width=1\textwidth]{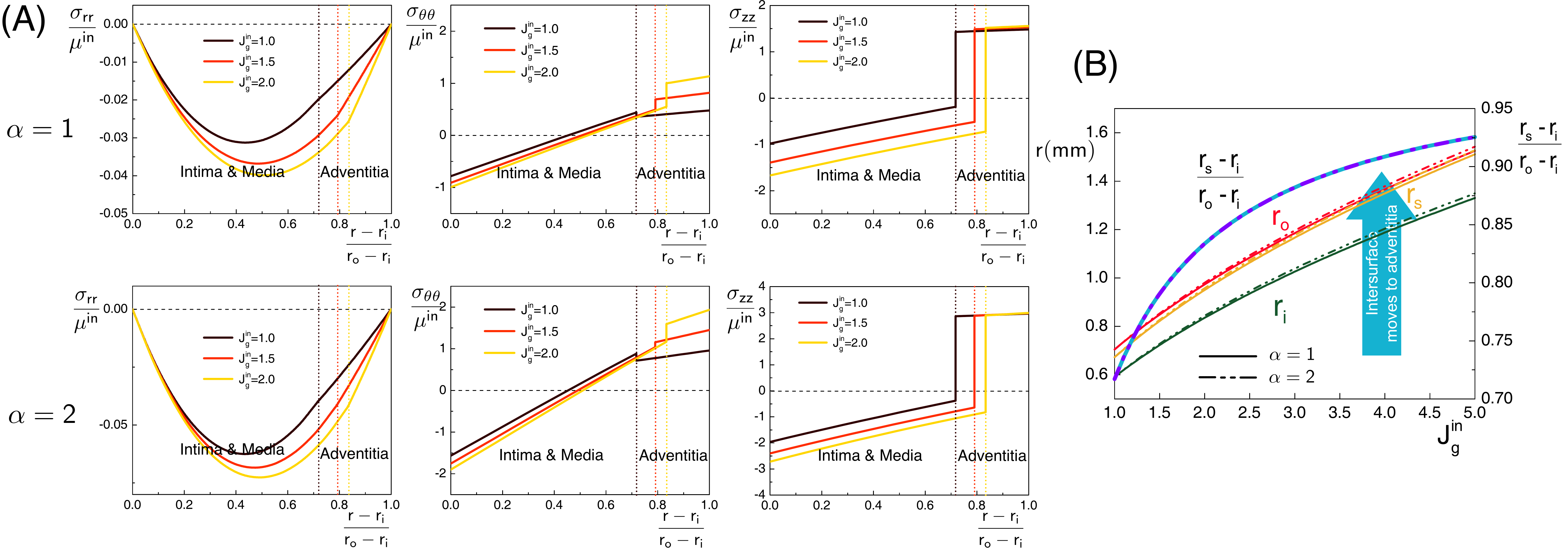}
	\caption{(A) Transmural distribution of growth-induced residual stress components in a bi-layered  aorta when the inner layer grows with different growth factors $J_g^{in}=1,1.5,2$, in the presence of an initial residual stress found in an aged human aorta\cite{Holzapfel2010, Holzapfel2007}, with magnitudes $\alpha=1$ and $\alpha=2$; (B) Changes in radii and thicknesses of the bi-layered aorta with increasing inner layer growth and different initial residual stress magnitudes $\alpha=1$ and $\alpha=2$.}
  	\label{RSGC}
\end{figure*}

Figure \ref{RSGC}(A) shows the transmural stress distributions in the aorta. 
Here the growth-induced residual stress follow\color{black}s \color{black} from growth factors $J_g^{in}=1,1.5,2$ in the inner layer, \color{black} $J_g^{ad}=1$ in adventitia layer\color{black}, and different initial residual stress magnitudes $\alpha=1$ (same as Holzapfel and collaborators\cite{Holzapfel2010, Holzapfel2007}) and $\alpha=2$ (twice their level). 
We see that the magnitude of the residual stress components increases with increasing growth volume in the inner layer. 
Moreover, the changes of residual stresses due to growth are very similar between the two different initial residual stress levels $\alpha=1$ and $\alpha=2$, which demonstrates that the level of self-balanced initial residual stress does not affect the accumulation law of growth-induced residual stress. 
However, owing to the superposition effect of initial residual stress and growth-induced residual stress, there will be a significant difference in the total residual stress. 

Figure \ref{RSGC}(B) shows the changes in radii and thicknesses of this bi-layered aorta with increasing growth of the inner layer ($1.0 \le J_g^{in} \le 5.0$) and different initial residual stress \color{black} levels \color{black} ($\alpha=1,2$). All the radii and the ratio of the inner layer's thickness to the adventitia's thickness $\frac{r_s-r_i}{r_o-r_s}$ are increasing with the isotropic growth of \color{black} the \color{black} inner layer, which is expected  due to the increasing volume of \color{black} the \color{black} inner layer. 
Moreover, the magnitude of the initial residual stress displays no effects on the thickness ratio of inner layer to adventitia $\frac{r_s-r_i}{r_o-r_s}$ but has some slight impact on the growth process by the little difference of the absolute radii.
This difference  could also play a role in creating wrinkles and patterns, see \color{black} the \color{black} next section.

We expect that increasing compressible circumferential stress and axial stress on the inner surface eventually induces a buckling of the surface and creates wrinkles or folds, a phenomenon that is commonly found in most tubular tissues. 
However, we see from Figure \ref{RSGC} that the circumferential tensile stress in the inner layer (intima and media) and the tensile axial stress in the outer layer (adventitia) are increased, which could have either opposite effects on instability or help the tissues keep stable.

Based on the MD growth model, Ciarletta et al. \cite{Ciarletta2014} have shown that  geometrical dimensions and stiffness contrast between the layers play a significant role in growth-induced pattern selection for tubular tissues. 
The results above have demonstrated  the influence of the initial residual stress on growth-induced residual stress accumulation and on geometry changes.
It follows that the initial residual stress must have a certain impact on the subsequent pattern creation, and  we now turn to wrinkling analysis to determine its exact influence . 

\section{Directional pattern creation by initial residual stress and differential growth}


Morphogenesis and pattern formation are closely related to several specific bio-functions. 
It is thus imperative to study the relationships between  growth, initial stress, change in shape and eventually, pattern selection.

Some pioneering works studied pattern selection for tubular tissues, where the tubular organ grows from a stress-free initial state. 
Hence Ciarletta et al. \cite{Ciarletta2014} showed that increasing the thickness or stiffness ratio between the outer and inner tubular layers creates fewer wrinkles and folds in the circumferential direction but more wrinkles in the axial direction, thus providing an insight into bio-medical engineering applications to achieve directional control or selection of growth-induced pattern creation. 
Here by setting $\alpha=0$, we expect to recover their results. 
Our innovation is a proposal to directionally control or select  patterns by initial residual stress distribution ($\alpha \ne 0$), instead of altering the geometry or elasticity of the tissues. 

We describe the wrinkles by a three-dimensional incremental field $\dot{\boldsymbol{x}}=u\left(r,\theta, z\right) \boldsymbol{e}_r+v\left(r,\theta, z\right) \boldsymbol{e}_\theta+w\left(r,\theta, z\right)\boldsymbol{e}_z$, (and $\dot p$, the increment of $p^k$), and seek solutions in the neighbourhood of the large deformation which have sinusoidal variations with $\theta$ and $z$, and can thus possibly form 2D-patterns. 
Hence we take
\begin{equation}
\left[u,\dot{p} \right]=\left[U(r),Q(r) \right] \cos(m\theta) \cos(\kappa z), \quad
\left[v, w\right]=\left[V(r), W(r)\right]\sin(m\theta) \cos(\kappa z),
\label{increments}
\end{equation}
where  $m$ and $\kappa=2\pi n/\ell$ are the circumferential and longitudinal wave numbers, respectively ($\ell$ is the current length of the tube). 
In other words, $m$ and $n$ are the numbers of  circumferential and axial wrinkles,  respectively.
Then the amplitude functions $U,V,W,Q$ are determined by solving numerically the incremental equilibrium equations and boundary conditions. 
This task is best achieved by using the Stroh formulation, which we recall in the appendix.
Then, by iterating over the wave numbers $m$ and $k$, we eventually get the critical value of each case which relates to the growth-induced pattern creation. 

\begin{figure}[h]
     \centering
    \includegraphics[width=1\textwidth]{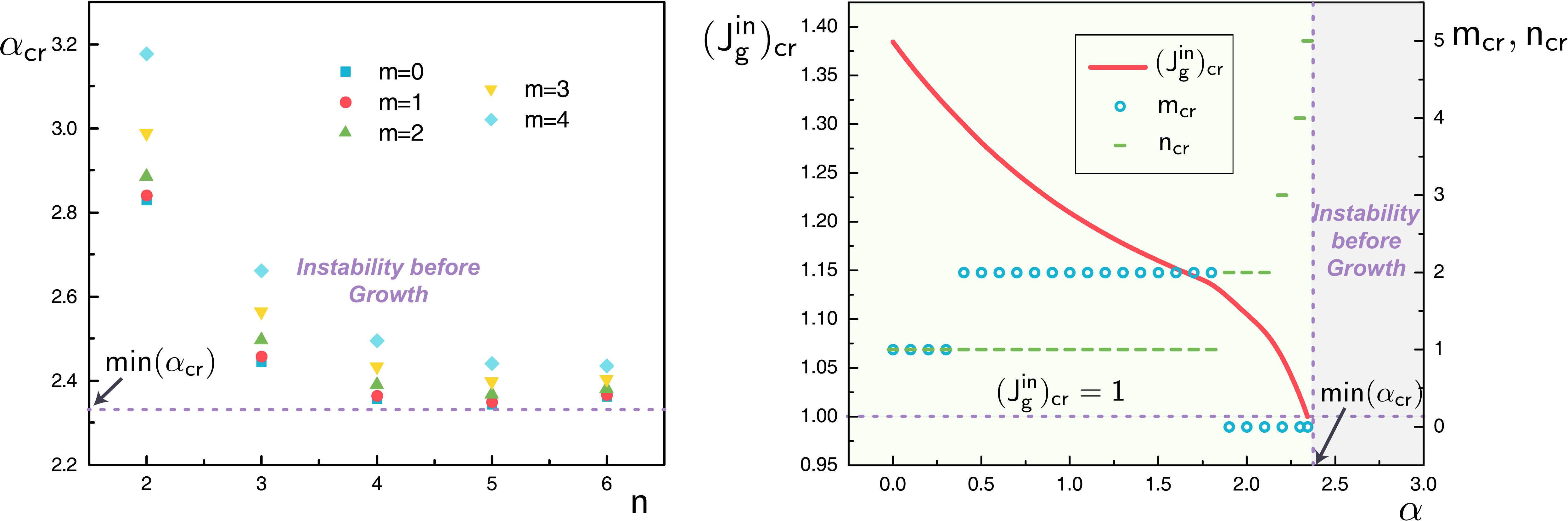}
    \caption{(A) Computation of the critical magnitude  $\alpha_\text{cr}$ of the initial residual stress in the aorta for patterns with $m$ cir\-cum\-fe\-ren\-tial wrinkles and $n$ axial wrinkles; (B) Influence of the magnitude of the initial residual stress (tracked by  $\alpha$) on inner layer growth-induced (tracked by $J_g^{in}$)  pattern creation (circles: number of circumferential wrinkles, dashes: number of axial wrinkles.)}
    \label{CA}
\end{figure}

First we need to evaluate the critical level of initial residual stress signalling when the aorta is unstable in its reference configuration before any differential growth takes place (for other examples of this situation, see\cite{Ciarletta2016, ciarletta2016residual,riccobelli2018shape}).
Here the initial residual stress is based on experimental data  and we quantify its influence by the magnitude factor $\alpha$. 
Since there are no wrinkling patterns of note in an aged human aorta, we must check that the initial residual state of Holzapfel et al.\cite{Holzapfel2010, Holzapfel2007} at $\alpha=1.0$ is in the stability range for $\alpha$.
Figure \ref{CA}(A) displays how the critical magnitude value $\alpha_\text{cr}$ relates to the possible wrinkling modes, with  number of circumferential wrinkles $m = 0,1,\ldots,4$ and number of axial wrinkles $n=1,2,\ldots, 6$.
The figure shows that the smallest $\alpha_\text{cr}$ for any given number of axial wrinkles $n$ occurs when there are no circumferential wrinkles ($m=0$): hence based only on an increase of initial residual stress (and no growth), the bi-layered aorta would buckle in the axial direction. Moreover, we also find the minimal critical value $\text{min}(\alpha_\text{cr}) = 2.343$, happening with $m=0$, $n=5$. 
It means that the surfaces of the initially residually stressed aorta (at $\alpha=1.0$)  are smooth and free of wrinkles according to this specific modeling, in line with experimental observation.

\begin{figure}
	 \centering
	\includegraphics[width=1\textwidth]{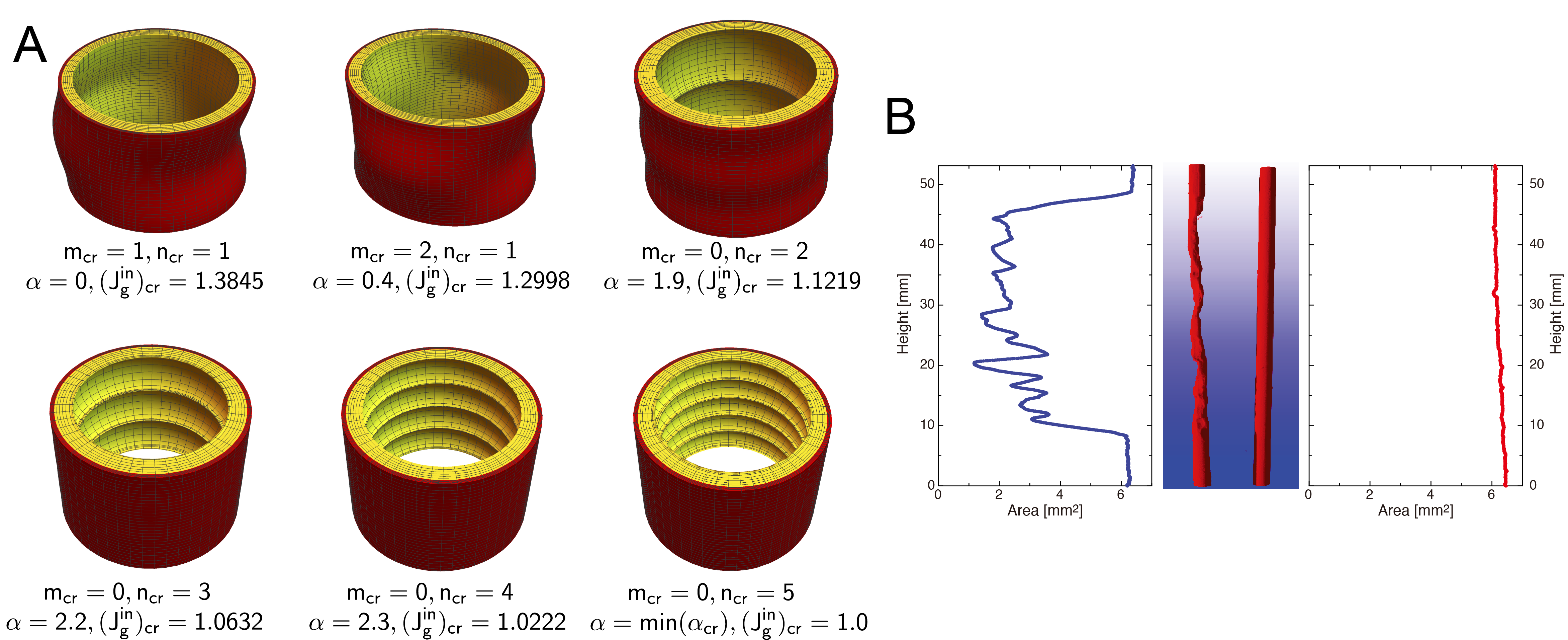}
	\caption{A. Typical wrinkle creations induced by growth of the inner layer ($J_g^{in}>0$) starting from certain initial residual stresses (as tracked by $\alpha$).
	When the numbers of circumferential ($m$) and axial ($n$) wrinkles are both non-zero, a 2D-pattern emerges, as in the first two examples here. 
	\color{black}
	B. Cross-sectional area of the inner volume of an atherosclerotic human coronary artery (left) and healthy (right) artery model derived from laboratory $\mu$CT measurements (from Holme et al.\cite{holme2012morphology}).}
	\color{black}
	\label{IIS}
\end{figure}

From now on we limit the span of magnitude $\alpha$ for the initial residual stress to be less than $\text{min}(\alpha_\text{cr})$, to avoid wrinkles in the initial configuration. 
Figure \ref{CA}(B) then shows the influence of $\alpha$ on growth-induced pattern creation.
 As $\alpha$ increases toward $\text{min}(\alpha_\text{cr})$, the amount of differential growth (volume accumulation in the inner layer) decreases, tending to 1 (no growth) eventually, as expected. 
In that limit we recover $m_\text{cr}=0$, $n_\text{cr}=5$ of Figure \ref{CA}(A).

The theoretical and numerical analysis in the appendix reveals that  critical pattern modes strongly depend on \emph{both} the magnitude $\alpha$ of the initial residual stress and the subsequent differential growth, rather than only on the latter as in previous studies. 
Figure \ref{CA} shows that growth-induced wrinkle creation is significantly altered by changing slightly the magnitude of initial residual stress and then letting the aorta grow at a constant rate. 
With no growth ($J_g^{in}$=0), the initially stress\color{black}ed \color{black} aorta can only wrinkle axially. 
With no initial residual stress ($\alpha=0$), differential growth can only lead to a $m=n=1$ pattern. 
By combining both effects, we can see circumferential and axial wrinkles arise and add constructively to form a further 2D pattern with $m=2$, $n=1$, see Figure \ref{IIS}(A) for examples.
\color{black} 
 Figure \ref{IIS}(B) shows a comparison of the cross-section areas of the inner volume between an atherosclerotic and a healthy aorta \cite{holme2012morphology}, which is qualitatively consistent with the conclusion drawn from Figure \ref{IIS}(A) when the initial residual stress gets bigger. So our results may provide another possible explanation of physical pathology for atherosclerosis.
\color{black} 
%


\section{Concluding remarks}


We demonstrate the influence of initial residual stress on the growth and pattern creation of layered aortas. 
A modified multiplicative decomposition growth (MMDG) framework is employed to incorporate and consider initial residual stress when we simulate the growth process theoretically.
We highlight the differences in residual stress accumulation and in geometrical changes due to the presence of initial residual stress. 
We also implement a linearized stability analysis to unveil pattern creation at critical levels of growth and initial stress. 

As a conclusion, we point out that understanding the effect of the initial residual stress provides an alternative way to select objective patterns or to control some bio-functions via morphological change. 
This may also present an inspiring insight for directional bionic self-assembly or robot manufacturing by initial residual stress. 


\section*{Acknowledgements}


We gratefully acknowledge the \color{black} support \color{black} from the National Natural Science Foundation of China through grants No.11621062 and No.11772295, as well as from the China Scholarship Council.
MD thanks Zhejiang University for support during visits to Hangzhou, and Annette Harte (Galway) for access to Irish Ash trees.
WQ also acknowledges the support from the Shenzhen Science, Technology and Innovation Commission for R \& D (No. JCYJ20170816172316775).


\section*{Author contributions statement}

Chaofeng L\"{u} and Yangkun Du conceived the main idea of the paper. 
Chaofeng L\"{u} and Yangkun Du developed the theory and performed the computations.
Michel Destrade and Weiqiu Chen verified the analytical methods. Chaofeng L\"{u} and Weiqiu Chen encouraged Yangkun Du to investigate the influence of initial residual stress on the pattern creation and supervised the findings of this work. Michel Destrade helped Yangkun Du with the wrinkling analysis part. All authors discussed the results and contributed to the final manuscript.


\section*{Competing interests statement}

The authors declare no competing interests.

\section*{Method}


We did not conduct ourselves the experiments on human aortas reported here. 
They were conducted by the authors of Reference [1] in 2007, and we are simply reporting some of their findings and using them as a base to derive some of our results. 
The use of autopsy material from human subjects in Reference [1] was approved in 2007 by the Ethics Committee, Medical University Graz, Austria, see details in Reference [1].


\section*{Appendix: Wrinkling analysis}


Here, we use an infinitesimal elastic deformation  $\chi'$ as a perturbation of the large deformation solution in the current configuration. 
The corresponding incremental displacement gradients  $\dot{\boldsymbol{F}}$ (with respect to the reference configuration)  and $\boldsymbol{\dot {F}}_I$ (with respect to the current configuration) are related as: $\dot{\boldsymbol{F}}=\dot{\boldsymbol{F}_I}\boldsymbol{F}$. 
By assuming that the incremental deformation is infinitesimal and transient, and that the corresponding growth process is independent of the stress and strain fields, we have the relationship: $\dot{\boldsymbol{F}_e}=\dot{\boldsymbol{F}_I}\boldsymbol{F}_e$ for the pure growth gradient.

With the incremental displacement field $\dot{\boldsymbol{x}}=u\left(r,\theta, z\right) \boldsymbol{e}_r+v\left(r,\theta, z\right) \boldsymbol{e}_\theta+w\left(r,\theta, z\right)\boldsymbol{e}_z$, we find 
the incremental displacement gradient tensor as
\begin{equation}
	\dot{\boldsymbol{F}_I}=\frac{\partial{\dot{\boldsymbol{x}}}}{\partial{\boldsymbol{x}}}=
	\begin{bmatrix}
		\dfrac{\partial{u}}{\partial{r}} &\dfrac{1}{r}\left(\dfrac{\partial u}{\partial \theta}-v\right) &\dfrac{\partial{u}}{\partial{r}} 	\\[12pt]
		\dfrac{\partial{v}}{\partial{r}} &\dfrac{1}{r}\left(\dfrac{\partial v}{\partial \theta}+u\right) &\dfrac{\partial{v}}{\partial{r}} 	\\[12pt]
		\dfrac{\partial{w}}{\partial{r}} &\dfrac{1}{r}\dfrac{\partial w}{\partial \theta} &\dfrac{\partial{w}}{\partial{z}}
	\end{bmatrix}
\end{equation}

The incremental nominal stress component in push-forward form is 
\begin{equation}
	\dot{{S}}_{Iij}=A_{eijkl}^I \dot{{F}}_{Ilk}-\dot{p}\delta_{ij}+p\dot{{F}}_{Iij},
\end{equation}
where $A_{eijkl}^I=F_{ei\alpha} F_{ek\beta} \frac{\partial \psi}{\partial F_{ej\alpha}\partial{F_{el\beta}}}$ are the instantaneous elastic moduli \cite{shams2011initial}.
%
The non-zero incremental equilibrium equations are along the three principal directions:
\begin{equation}
	\begin{split}
		&\left(\mathrm{div}\dot{\boldsymbol{S}}_I\right)_r=\frac{\partial{\dot{S}_{Irr}}}{\partial r}+\frac{\partial{\dot{S}_{I\theta r}}}{r \partial \theta}+\frac{\dot{S}_{Irr}-\dot{S}_{I\theta\theta}}{r}+\frac{\dot{\partial{S}_{Izr}}}{\partial z}=0,\\
		&\left(\mathrm{div}\dot{\boldsymbol{S}}_I\right)_\theta=\frac{\partial{\dot{S}_{Ir\theta}}}{\partial r}+\frac{\partial{\dot{S}_{I\theta \theta}}}{r \partial \theta}+\frac{\dot{S}_{Ir\theta}+\dot{S}_{I\theta r}}{r}+\frac{\dot{\partial{S}_{Iz\theta}}}{\partial z}=0,\\
		&\left(\mathrm{div}\dot{\boldsymbol{S}}_I\right)_z=\frac{\partial{\dot{S}_{Irz}}}{\partial r}+\frac{\partial{\dot{S}_{I\theta z}}}{r \partial \theta}+\frac{\dot{S}_{Irz}}{r}+\frac{\dot{\partial{S}_{Izz}}}{\partial z}=0
	\end{split}
	\label{incremental equilibrium equation}
\end{equation}
together with the incremental incompressibility condition,
\begin{equation}
	\mathrm{tr}\dot{\boldsymbol{F}_I}=\frac{\partial u}{\partial r}+\frac{1}{r}\left(\frac{\partial v}{\partial \theta}+u\right)+\frac{\partial{w}}{\partial{z}}=0.
	\label{incompressible condition}
\end{equation}
\color{black}
Using the forms \eqref{increments} for the incremental fields and the corresponding form of the incremental nominal stress in push-forward form
\begin{equation}
	\begin{split}
		&\dot{S}_{Irr}\left(r,\theta,z\right)=\Sigma_{rr} \left(r\right) \cos \left(m\theta\right) \cos \left(\kappa z\right),\qquad\dot{S}_{Ir\theta}\left(r,\theta,z\right)=\Sigma_{r\theta} \left(r\right) \sin \left(m\theta\right) \cos \left(\kappa z\right),\\		&\dot{S}_{Irz}\left(r,\theta,z\right)=\Sigma_{rz} \left(r\right) \cos \left(m\theta\right) \sin \left(\kappa z\right).
	\end{split}
	\label{increments2}
\end{equation}
we are able to re-organize the governing equations in the Stroh form as
\color{black}
\begin{equation} 
	\frac{d\boldsymbol{\eta}(r)}{dr}=\frac{1}{r}
	\begin{bmatrix} 
	\boldsymbol{G}_1(r) & \boldsymbol{G}_2(r)\\
	\boldsymbol{G}_3(r) & -\boldsymbol{G}_1^T(r)
	\end{bmatrix}
	\boldsymbol{\eta}(r)
	\label{Stroh}
\end{equation}
where $\boldsymbol{\eta}(r)=[U(r), V(r), W(r), r\Sigma_{rr}(r), r\Sigma_{r\theta}(r), r\Sigma_{rz}(r)]^T$ and 
\begin{align*}
& \boldsymbol{G}_1(r)=
	\begin{bmatrix} 
		-1 & -m & -\kappa r \\
		\dfrac{m(A_{e r \theta r \theta}^I-\sigma_{rr})}{A_{e r \theta r \theta}^I} &\dfrac{(A_{e r \theta r \theta}^I-\sigma_{rr})}{A_{e r \theta r \theta}^I} & 0\\
		\dfrac{\kappa r(A_{e rzrz^I-\sigma_{rr})}}{A_{erzrz}^I} & 0 & 0\\
	\end{bmatrix}, 
	\\
	&
\boldsymbol{G}_2(r)=
	\begin{bmatrix} 
		0& 0& 0 \\
		0 &\dfrac{1}{A_{er\theta r\theta}^I} &0\\
		0 &0 &\dfrac{1}{A_{erzrz}^I} 
	\end{bmatrix},
\\
&
\boldsymbol{G}_3(r)=
\begin{bmatrix}
	k_{11} &k_{12} &k_{13}\\
	k_{21} &k_{22} &k_{23}\\
	k_{13} &k_{23} &k_{33}\\
\end{bmatrix}, \notag \\
&
k_{11}=m^2\left(A_{e\theta r \theta r}^I-\dfrac{(A_{er \theta r \theta}^I-\sigma_{rr})^2}{A_{er \theta r \theta}^I}\right)+A_{e\theta\theta \theta \theta}^I+2p
\\
& \qquad\qquad +\kappa^2r^2\left(A_{ezrzr}^I-\dfrac{(A_{erzrz}^I-\sigma_{rr})^2}{A_{erzrz}^I}\right)+A_{errrr}^{I}, \notag \\
&k_{12}=m\left(A_{e\theta r\theta r}^I+A_{e\theta\theta\theta\theta}^I+2p-\dfrac{(A_{er \theta r \theta}^I-\sigma_{rr})^2}{A_{er \theta r \theta}^I}+A_{errrr}^{I}\right),\\
&k_{13}=\kappa r(A_{errrr}^{I}+p),\\
&k_{22}=m^2(A_{e\theta\theta\theta\theta}^{I}+2p+A_{errrr}^{I})+A_{e\theta r\theta r}^{I}+\kappa^2r^2A_{ez\theta z\theta}^{I}-\dfrac{(A_{er \theta r \theta}^I-\sigma_{rr})^2}{A_{er \theta r \theta}^I},\\
&k_{23}=\kappa mr(2p+A_{errrr}^I), k_{33}=A_{e\theta z\theta z}m^2+\kappa^2r^2(A_{ezzzz}^I+2p+A_{errrr}^I).
\end{align*} 

To solve these equations numerically, subject to the appropriate incremental boundary conditions, we use the surface impedance method, see details elsewhere\cite{Biryukov1985350, DESTRADE20101212, DESTRADE20094322}.


\begin{thebibliography}{10}

\expandafter\ifx\csname url\endcsname\relax
  \def\url#1{\texttt{#1}}\fi
\expandafter\ifx\csname urlprefix\endcsname\relax\def\urlprefix{URL }\fi
\expandafter\ifx\csname doiprefix\endcsname\relax\def\doiprefix{DOI: }\fi
\providecommand{\bibinfo}[2]{#2}
\providecommand{\eprint}[2][]{\url{#2}}

\bibitem{Holzapfel2007}
\bibinfo{author}{Holzapfel, G.~A.}, \bibinfo{author}{Sommer, G.},
  \bibinfo{author}{Auer, M.}, \bibinfo{author}{Regitnig, P.} \&
  \bibinfo{author}{Ogden, R.~W.}
\newblock \bibinfo{journal}{\bibinfo{title}{{Layer-specific 3D residual
  deformations of human aortas with non-atherosclerotic intimal thickening}}}.
\newblock {\emph{{Annals of Biomedical Engineering}}}
  \textbf{\bibinfo{volume}{35}}, \bibinfo{pages}{530--545},
  \doiprefix\url{10.1007/s10439-006-9252-z} (\bibinfo{year}{2007}).

\bibitem{Holzapfel2010}
\bibinfo{author}{Holzapfel, G.~A.} \& \bibinfo{author}{Ogden, R.~W.}
\newblock \bibinfo{journal}{\bibinfo{title}{{Modelling the layer-specific
  three-dimensional residual stresses in arteries, with an application to the
  human aorta}}}.
\newblock {\emph{{Journal of the Royal Society Interface}}}
  \textbf{\bibinfo{volume}{7}}, \bibinfo{pages}{787--799},
  \doiprefix\url{10.1098/rsif.2009.0357} (\bibinfo{year}{2010}).

\bibitem{Omens1990}
\bibinfo{author}{Omens, J.~H.} \& \bibinfo{author}{Fung, Y.-C.}
\newblock \bibinfo{journal}{\bibinfo{title}{{Residual strain in rat left
  ventricle}}}.
\newblock {\emph{{Circulation Research}}}
  \textbf{\bibinfo{volume}{66}}, \bibinfo{pages}{37--45}
  (\bibinfo{year}{1990}).

\bibitem{Savin2011}
\bibinfo{author}{Savin, T.} \emph{et~al.}
\newblock \bibinfo{journal}{\bibinfo{title}{{On the growth and form of the
  gut}}}.
\newblock {\emph{{Nature}}} \textbf{\bibinfo{volume}{476}},
  \bibinfo{pages}{57--62}, \doiprefix\url{10.1038/nature10277}
  (\bibinfo{year}{2011}).

\bibitem{Balbi2015}
\bibinfo{author}{Balbi, V.}, \bibinfo{author}{Kuhl, E.} \&
  \bibinfo{author}{Ciarletta, P.}
\newblock \bibinfo{journal}{\bibinfo{title}{{Morphoelastic control of
  gastro-intestinal organogenesis: Theoretical predictions and numerical
  insights}}}.
\newblock {\emph{{Journal of the Mechanics and Physics of
  Solids}}} \textbf{\bibinfo{volume}{78}}, \bibinfo{pages}{493--510},
  \doiprefix\url{10.1016/j.jmps.2015.02.016} (\bibinfo{year}{2015}).

\bibitem{Stylianopoulos2012}
\bibinfo{author}{Stylianopoulos, T.} \emph{et~al.}
\newblock \bibinfo{journal}{\bibinfo{title}{{Causes, consequences, and remedies
  for growth-induced solid stress in murine and human tumors}}}.
\newblock {\emph{{Proceedings of the National Academy of
  Sciences}}} \textbf{\bibinfo{volume}{109}}, \bibinfo{pages}{15101--15108},
  \doiprefix\url{10.1073/pnas.1213353109} (\bibinfo{year}{2012}).

\bibitem{Fernandez-Sanchez2015}
\bibinfo{author}{Fernandez-Sanchez, M.~E.} \emph{et~al.}
\newblock \bibinfo{journal}{\bibinfo{title}{{Mechanical induction of the
  tumorigenic beta-catenin pathway by tumour growth pressure}}}.
\newblock {\emph{{Nature}}} \textbf{\bibinfo{volume}{523}},
  \bibinfo{pages}{92--95}, \doiprefix\url{10.1038/nature14329}
  (\bibinfo{year}{2015}).

\bibitem{Helmlinger1997}
\bibinfo{author}{Helmlinger, G.}, \bibinfo{author}{Netti, P.~A.},
  \bibinfo{author}{Lichtenbeld, H.~C.}, \bibinfo{author}{Melder, R.~J.} \&
  \bibinfo{author}{Jain, R.~K.}
\newblock \bibinfo{journal}{\bibinfo{title}{{Solid stress inhibits the growth
  of multicellular tumor spheroids}}}.
\newblock {\emph{{Nature Biotechnology}}}
  \textbf{\bibinfo{volume}{15}}, \bibinfo{pages}{778} (\bibinfo{year}{1997}).

\bibitem{Padera2004}
\bibinfo{author}{Padera, T.~P.} \emph{et~al.}
\newblock \bibinfo{journal}{\bibinfo{title}{{Pathology: Cancer cells compress
  intratumour vessels}}}.
\newblock {\emph{{Nature}}} \textbf{\bibinfo{volume}{427}},
  \bibinfo{pages}{695} (\bibinfo{year}{2004}).

\bibitem{destrade2012uniform}
\bibinfo{author}{Destrade, M.}, \bibinfo{author}{Liu, Y.},
  \bibinfo{author}{Murphy, J.~G.} \& \bibinfo{author}{Kassab, G.~S.}
\newblock \bibinfo{journal}{\bibinfo{title}{Uniform transmural strain in
  pre-stressed arteries occurs at physiological pressure}}.
\newblock {\emph{{Journal of Theoretical Biology}}}
  \textbf{\bibinfo{volume}{303}}, \bibinfo{pages}{93--97}
  (\bibinfo{year}{2012}).

\bibitem{Pan2016}
\bibinfo{author}{Pan, Y.}, \bibinfo{author}{Heemskerk, I.},
  \bibinfo{author}{Ibar, C.}, \bibinfo{author}{Shraiman, B.~I.} \&
  \bibinfo{author}{Irvine, K.~D.}
\newblock \bibinfo{journal}{\bibinfo{title}{{Differential growth triggers
  mechanical feedback that elevates Hippo signaling}}}.
\newblock {\emph{{Proceedings of the National Academy of
  Sciences}}} \bibinfo{pages}{201615012} (\bibinfo{year}{2016}).

\bibitem{Rodriguez1994}
\bibinfo{author}{Rodriguez, E.~K.}, \bibinfo{author}{Hoger, A.} \&
  \bibinfo{author}{Mcculloch, A.~D.}
\newblock \bibinfo{journal}{\bibinfo{title}{{Stress-dependent finite growth in
  soft elastic tissues}}}.
\newblock {\emph{{Journal of Biomechanics}}}
  \textbf{\bibinfo{volume}{27}}, \bibinfo{pages}{455--467},
  \doiprefix\url{10.1016/0021-9290(94)90021-3} (\bibinfo{year}{1994}).

\bibitem{Cao2012}
\bibinfo{author}{Cao, Y.-P.}, \bibinfo{author}{Li, B.} \&
  \bibinfo{author}{Feng, X.-Q.}
\newblock \bibinfo{journal}{\bibinfo{title}{{Surface wrinkling and folding of
  core--shell soft cylinders}}}.
\newblock {\emph{{Soft Matter}}} \textbf{\bibinfo{volume}{8}},
  \bibinfo{pages}{556--562}, \doiprefix\url{10.1039/c1sm06354e}
  (\bibinfo{year}{2012}).

\bibitem{Li2011}
\bibinfo{author}{Li, B.}, \bibinfo{author}{Cao, Y.-P.}, \bibinfo{author}{Feng,
  X.-Q.} \& \bibinfo{author}{Gao, H.}
\newblock \bibinfo{journal}{\bibinfo{title}{{Surface wrinkling of mucosa
  induced by volumetric growth: Theory, simulation and experiment}}}.
\newblock {\emph{{Journal of the Mechanics and Physics of
  Solids}}} \textbf{\bibinfo{volume}{59}}, \bibinfo{pages}{758--774},
  \doiprefix\url{10.1016/j.jmps.2011.01.010} (\bibinfo{year}{2011}).

\bibitem{Lu2016}
\bibinfo{author}{L{\"{u}}, C.~F.} \& \bibinfo{author}{Du, Y.~K.}
\newblock \bibinfo{journal}{\bibinfo{title}{{Theoretical modeling for
  monitoring the growth of fusiform abdominal aortic aneurysms using dielectric
  elastomer capacitive sensors}}}.
\newblock {\emph{{International Journal of Applied Mechanics}}}
  \textbf{\bibinfo{volume}{08}}, \bibinfo{pages}{1640010},
  \doiprefix\url{10.1142/S175882511640010X} (\bibinfo{year}{2016}).

\bibitem{Du2017}
\bibinfo{author}{Du, Y.} \& \bibinfo{author}{L{\"{u}}, C.}
\newblock \bibinfo{journal}{\bibinfo{title}{{Modeling on monitoring the growth
  and rupture assessment of saccular aneurysms}}}.
\newblock {\emph{{Theoretical and Applied Mechanics Letters}}}
  \textbf{\bibinfo{volume}{7}}, \bibinfo{pages}{117--120},
  \doiprefix\url{10.1016/j.taml.2017.01.007} (\bibinfo{year}{2017}).

\bibitem{Wang2017}
\bibinfo{author}{Wang, Y.~Z.}, \bibinfo{author}{Zhang, C.~L.} \&
  \bibinfo{author}{Chen, W.~Q.}
\newblock \bibinfo{journal}{\bibinfo{title}{An analytical model to predict
  material gradient and anisotropy in bamboo}}.
\newblock {\emph{{Acta Mechanica}}}
  \textbf{\bibinfo{volume}{228}}, \bibinfo{pages}{2819--2833},
  \doiprefix\url{10.1007/s00707-015-1514-0} (\bibinfo{year}{2017}).

\bibitem{Goriely2005}
\bibinfo{author}{Goriely, A.} \& \bibinfo{author}{{Ben Amar}, M.}
\newblock \bibinfo{journal}{\bibinfo{title}{{Differential growth and
  instability in elastic shells}}}.
\newblock {\emph{{Physical Review Letters}}}
  \textbf{\bibinfo{volume}{94}}, \bibinfo{pages}{198103},
  \doiprefix\url{10.1103/PhysRevLett.94.198103} (\bibinfo{year}{2005}).

\bibitem{Ciarletta2013}
\bibinfo{author}{Ciarletta, P.}
\newblock \bibinfo{journal}{\bibinfo{title}{{Buckling instability in growing
  tumor spheroids}}}.
\newblock {\emph{{Physical Review Letters}}}
  \textbf{\bibinfo{volume}{110}}, \bibinfo{pages}{158102},
  \doiprefix\url{10.1103/PhysRevLett.110.158102} (\bibinfo{year}{2013}).

\bibitem{Ciarletta2014}
\bibinfo{author}{Ciarletta, P.}, \bibinfo{author}{Balbi, V.} \&
  \bibinfo{author}{Kuhl, E.}
\newblock \bibinfo{journal}{\bibinfo{title}{{Pattern selection in growing
  tubular tissues}}}.
\newblock {\emph{{Physical Review Letters}}}
  \textbf{\bibinfo{volume}{113}}, \bibinfo{pages}{248101},
  \doiprefix\url{10.1103/PhysRevLett.113.248101} (\bibinfo{year}{2014}).

\bibitem{Johnson1995}
\bibinfo{author}{Johnson, B.~E.} \& \bibinfo{author}{Hoger, A.}
\newblock \bibinfo{journal}{\bibinfo{title}{{The use of a virtual configuration
  in formulating constitutive equations for residually stressed elastic
  materials}}}.
\newblock {\emph{{Journal of Elasticity}}}
  \textbf{\bibinfo{volume}{41}}, \bibinfo{pages}{177--215},
  \doiprefix\url{10.1007/bf00041874} (\bibinfo{year}{1995}).

\bibitem{Goriely2007}
\bibinfo{author}{Goriely, A.} \& \bibinfo{author}{{Ben Amar}, M.}
\newblock \bibinfo{journal}{\bibinfo{title}{{On the definition and modeling of
  incremental, cumulative, and continuous growth laws in morphoelasticity}}}.
\newblock {\emph{{Biomechanics and Modeling in Mechanobiology}}}
  \textbf{\bibinfo{volume}{6}}, \bibinfo{pages}{289--296},
  \doiprefix\url{10.1007/s10237-006-0065-7} (\bibinfo{year}{2007}).

\bibitem{Ciarletta2016}
\bibinfo{author}{Ciarletta, P.}, \bibinfo{author}{Destrade, M.},
  \bibinfo{author}{Gower, A.~L.} \& \bibinfo{author}{Taffetani, M.}
\newblock \bibinfo{journal}{\bibinfo{title}{{Morphology of residually stressed
  tubular tissues: Beyond the elastic multiplicative decomposition}}}.
\newblock {\emph{{Journal of the Mechanics and Physics of
  Solids}}} \textbf{\bibinfo{volume}{90}}, \bibinfo{pages}{242--253},
  \doiprefix\url{10.1016/j.jmps.2016.02.020} (\bibinfo{year}{2016}).

\bibitem{Hoger1986}
\bibinfo{author}{Hoger, A.}
\newblock \bibinfo{journal}{\bibinfo{title}{{On the determination of residual
  stress in an elastic body}}}.
\newblock {\emph{{Journal of Elasticity}}}
  \textbf{\bibinfo{volume}{16}}, \bibinfo{pages}{303--324}
  (\bibinfo{year}{1986}).

\bibitem{Shams2011}
\bibinfo{author}{Shams, M.}, \bibinfo{author}{Destrade, M.} \&
  \bibinfo{author}{Ogden, R.~W.}
\newblock \bibinfo{journal}{\bibinfo{title}{{Initial stresses in elastic
  solids: Constitutive laws and acoustoelasticity}}}.
\newblock {\emph{{Wave Motion}}} \textbf{\bibinfo{volume}{48}},
  \bibinfo{pages}{552--567}, \doiprefix\url{10.1016/j.wavemoti.2011.04.004}
  (\bibinfo{year}{2011}).

\bibitem{Gower2015}
\bibinfo{author}{Gower, A.~L.}, \bibinfo{author}{Ciarletta, P.} \&
  \bibinfo{author}{Destrade, M.}
\newblock \bibinfo{journal}{\bibinfo{title}{{Initial stress symmetry and its
  applications in elasticity}}}.
\newblock {\emph{{Proceedings of the Royal Society A}}}
  \textbf{\bibinfo{volume}{471}}, \bibinfo{pages}{20150448},
  \doiprefix\url{10.1098/rspa.2015.0448} (\bibinfo{year}{2015}).

\bibitem{Du2018}
\bibinfo{author}{Du, Y.}, \bibinfo{author}{L{\"{u}}, C.},
  \bibinfo{author}{Chen, W.} \& \bibinfo{author}{Destrade, M.}
\newblock \bibinfo{journal}{\bibinfo{title}{{Modified multiplicative
  decomposition model for tissue growth: Beyond the initial stress-free
  state}}}.
\newblock {\emph{{Journal of the Mechanics and Physics of
  Solids}}} \textbf{\bibinfo{volume}{118}},
  \doiprefix\url{10.1016/j.jmps.2018.05.014} (\bibinfo{year}{2018}).

\bibitem{Yin2009}
\bibinfo{author}{Yin, J.}, \bibinfo{author}{Bar-Kochba, E.} \&
  \bibinfo{author}{Chen, X.}
\newblock \bibinfo{journal}{\bibinfo{title}{{Mechanical self-assembly
  fabrication of gears}}}.
\newblock {\emph{{Soft Matter}}} \textbf{\bibinfo{volume}{5}},
  \bibinfo{pages}{3469--3474}, \doiprefix\url{10.1039/b904635f}
  (\bibinfo{year}{2009}).

\bibitem{Chen2010}
\bibinfo{author}{Chen, X.} \& \bibinfo{author}{Yin, J.}
\newblock \bibinfo{journal}{\bibinfo{title}{{Buckling patterns of thin films on
  curved compliant substrates with applications to morphogenesis and
  three-dimensional micro-fabrication}}}.
\newblock {\emph{{Soft Matter}}} \textbf{\bibinfo{volume}{6}},
  \bibinfo{pages}{5667--5680} (\bibinfo{year}{2010}).

\bibitem{Chen2013}
\bibinfo{author}{Chen, X.}
\newblock \emph{\bibinfo{title}{{Mechanical Self-Assembly}}}
  (\bibinfo{year}{2013}).
\newblock \eprint{arXiv:1011.1669v3}.

\bibitem{Li2013}
\bibinfo{author}{Li, B.}, \bibinfo{author}{Xu, G.-K.} \& \bibinfo{author}{Feng,
  X.-Q.}
\newblock \bibinfo{journal}{\bibinfo{title}{{Tissue-growth model for the
  swelling analysis of core-shell hydrogels}}}.
\newblock {\emph{{Soft Materials}}} \textbf{\bibinfo{volume}{11}},
  \bibinfo{pages}{117--124}, \doiprefix\url{10.1080/1539445x.2011.584603}
  (\bibinfo{year}{2013}).

\bibitem{Bertrand2016}
\bibinfo{author}{Bertrand, T.}, \bibinfo{author}{Peixinho, J.},
  \bibinfo{author}{Mukhopadhyay, S.} \& \bibinfo{author}{MacMinn, C.~W.}
\newblock \bibinfo{journal}{\bibinfo{title}{{Dynamics of swelling and drying in
  a spherical gel}}}.
\newblock {\emph{{Physical Review Applied}}}
  \textbf{\bibinfo{volume}{6}}, \doiprefix\url{10.1103/PhysRevApplied.6.064010}
  (\bibinfo{year}{2016}).

\bibitem{merodio2013influence}
\bibinfo{author}{Merodio, J.}, \bibinfo{author}{Ogden, R.~W.} \&
  \bibinfo{author}{Rodr{\'\i}guez, J.}
\newblock \bibinfo{journal}{\bibinfo{title}{The influence of residual stress on
  finite deformation elastic response}}.
\newblock {\emph{{International Journal of Non-Linear Mechanics}}}
  \textbf{\bibinfo{volume}{56}}, \bibinfo{pages}{43--49}
  (\bibinfo{year}{2013}).

\bibitem{ahamed2016modelling}
\bibinfo{author}{Ahamed, T.}, \bibinfo{author}{Dorfmann, L.} \&
  \bibinfo{author}{Ogden, R.}
\newblock \bibinfo{journal}{\bibinfo{title}{Modelling of residually stressed
  materials with application to {AAA}}}.
\newblock {\emph{{Journal of the Mechanical Behavior of Biomedical
  Materials}}} \textbf{\bibinfo{volume}{61}}, \bibinfo{pages}{221--234}
  (\bibinfo{year}{2016}).

\bibitem{merodio2016extension}
\bibinfo{author}{Merodio, J.} \& \bibinfo{author}{Ogden, R.~W.}
\newblock \bibinfo{journal}{\bibinfo{title}{Extension, inflation and torsion of
  a residually stressed circular cylindrical tube}}.
\newblock {\emph{{Continuum Mechanics and Thermodynamics}}}
  \textbf{\bibinfo{volume}{28}}, \bibinfo{pages}{157--174}
  (\bibinfo{year}{2016}).

\bibitem{ciarletta2016residual}
\bibinfo{author}{Ciarletta, P.}, \bibinfo{author}{Destrade, M.} \&
  \bibinfo{author}{Gower, A.~L.}
\newblock \bibinfo{journal}{\bibinfo{title}{On residual stresses and
  homeostasis: an elastic theory of functional adaptation in living matter}}.
\newblock {\emph{{Scientific Reports}}}
  \textbf{\bibinfo{volume}{6}}, \bibinfo{pages}{24390} (\bibinfo{year}{2016}).

\bibitem{riccobelli2018shape}
\bibinfo{author}{Riccobelli, D.} \& \bibinfo{author}{Ciarletta, P.}
\newblock \bibinfo{journal}{\bibinfo{title}{Shape transitions in a soft
  incompressible sphere with residual stresses}}.
\newblock {\emph{{Mathematics and Mechanics of Solids}}}
  \textbf{\bibinfo{volume}{23}}, \bibinfo{pages}{1507--1524}
  (\bibinfo{year}{2018}).

\bibitem{holme2012morphology}
\bibinfo{author}{Holme, M.~N.} \emph{et~al.}
\newblock \bibinfo{title}{Morphology of atherosclerotic coronary arteries}.
\newblock In \emph{\bibinfo{booktitle}{Developments in X-Ray Tomography VIII}},
  vol. \bibinfo{volume}{8506}, \bibinfo{pages}{850609}
  (\bibinfo{organization}{International Society for Optics and Photonics},
  \bibinfo{year}{2012}).

\bibitem{Biryukov1985350}
\bibinfo{author}{Biryukov, S.~V.}
\newblock \bibinfo{journal}{\bibinfo{title}{Impedance method in the theory of
  elastic surface waves}}.
\newblock {\emph{{Sov. Phys. Acoust.}}}
  \textbf{\bibinfo{volume}{31}}, \bibinfo{pages}{350--354}
  (\bibinfo{year}{1985}).

\bibitem{DESTRADE20101212}
\bibinfo{author}{Destrade, M.}, \bibinfo{author}{Murphy, J.~G.} \&
  \bibinfo{author}{Ogden, R.~W.}
\newblock \bibinfo{journal}{\bibinfo{title}{On deforming a sector of a circular
  cylindrical tube into an intact tube: Existence, uniqueness, and stability}}.
\newblock {\emph{{International Journal of Engineering Science}}}
  \textbf{\bibinfo{volume}{48}}, \bibinfo{pages}{1212 -- 1224}
  (\bibinfo{year}{2010}).

\bibitem{DESTRADE20094322}
\bibinfo{author}{Destrade, M.}, \bibinfo{author}{Annaidh, A.~N.} \&
  \bibinfo{author}{Coman, C.~D.}
\newblock \bibinfo{journal}{\bibinfo{title}{Bending instabilities of soft
  biological tissues}}.
\newblock {\emph{{International Journal of Solids and
  Structures}}} \textbf{\bibinfo{volume}{46}}, \bibinfo{pages}{4322 -- 4330}
  (\bibinfo{year}{2009}).

\end{thebibliography}
\end{document}